\documentclass[11pt,twoside]{article}
\usepackage{./asp2010}
\usepackage{graphics}
\resetcounters

\begin{document}

\title{Sparse Aperture Masking of Massive Stars}
\author{Sana H.$^1$, Lacour S.$^2$, Le Bouquin J.-B.$^3$, de Koter A.$^{1,4}$, Moni Bidin C.$^5$, Muijres L.$^1$, Schnurr O.$^6$, Zinnecker H.$^{7,8}$
\affil{$^1$ Astronomical Institute `Anton Pannekoek', University of Amsterdam, Postbus 94249, 1090 GE, Amsterdam, The Netherlands}
\affil{$^2$ LESIA, Observatoire de Paris, Universit\'e Paris Diderot,  Meudon, France}
\affil{$^3$ Institut de Plan{\'e}tologie et d'Astrophysique de Grenoble (IPAG),  France}
\affil{$^4$ Astronomical Institute, Utrecht University, Utrecht, The Netherlands}
\affil{$^5$ Departamento de Astronom\'ia, Universidad de Concepci\'on, Chile}
\affil{$^6$ Leibniz-Institut für Astrophysik Potsdam (AIP), Potsdam, Germany}
\affil{$^7$ Deutsches SOFIA Institute, Stuttgart, Germany}
\affil{$^8$ SOFIA Science Center, NASA-Ames Research Center, Moffett Field, CA, USA}
}

\begin{abstract}
We present the earliest results of our NACO/VLT sparse aperture masking (SAM) campaign to search for binarity in a sample of 60 O-type stars.
We detect $\Delta K\mathrm{s} < 5$~mag companions for 20-25\%\ of our targets with separations in the range 30-100 mas (typically, 40 -
200 A.U.). Most of these companions were unknown, shedding thus new light on the multiplicity properties of massive
stars in a separation and brightness regime that has been difficult to explore so far. Adding detections
from other techniques (spectroscopy, interferometry, speckle, lucky imaging, AO), the fraction of O stars with at least one companion is  85\%\ (51/60
targets). This is the largest multiplicity fraction ever found.
\end{abstract}

\section{Introduction}

With masses $\ga$ 16 M$_\odot$, massive stars of spectral type O are among the brightest and most luminous stars in galaxies. One of their most striking properties is their high multiplicity rate \citep[for recent reviews, see ][ and Gies -- this volume]{SaE11}: above 40\% for visual systems \citep{TBR08,MHG09} and up to $\sim$60\% for spectroscopic binaries \citep{MHG09, SaE11}. In nearby clusters, at least 75\%\ of the massive stars are part of a binary or higher multiplicity systems \citep{SGN08, SGE09, SJG11}. The multiplicity fraction  and distribution of the binary parameters (periods, mass ratios, eccentricities) are one of the few observable quantities that can help constrain the formation and early dynamical evolution of these objects (see e.g., Kratter et al. this volume) and can potentially discriminate between the various scenarios \citep[see e.g.][for a review]{ZiY07}. However, the census of the properties of the massive star population remains incomplete.

\begin{figure}
  \includegraphics[width=11cm]{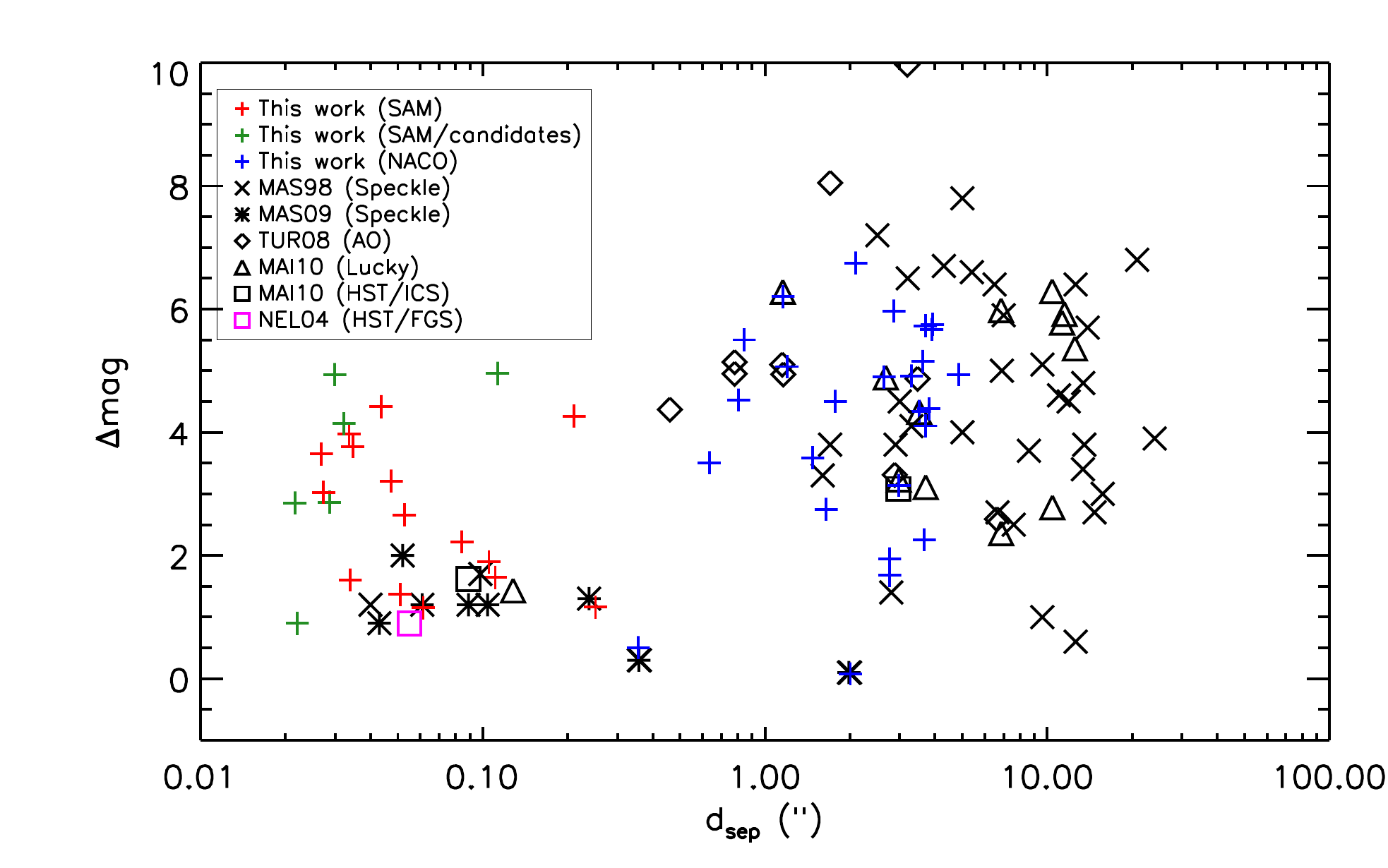}
  \caption{Magnitude difference vs. separation for all the pairs in our sample. Different symbols/colors indicate different observing techniques. Objects observed with various instruments have multiple entries in the figure, possibly with slightly different $\Delta$mag due to the different bands in which the techniques are operating. Using SAM, we are now able to probe the short separation/high contrast regime. References: MAS98 -- \citet{MGH98}, NEL04 -- \citet{NWW04}, TUR08 -- \citet{TBR08}, MAS09 -- \citet{MHG09}, MAI10 -- \citet{MAp10}.}
  \label{fig1}
\end{figure}


\section{Early results}

In March 2011, we observed a  sample of 60 O stars with $K<7.5$~mag with the SAM mode of NACO, providing an almost bias-free detection up to a flux contrast of 100 in the range 30-200~mas. Under the adopted configuration (512$\times$512 windowing), the NACO field of view  extends over 6''$\times$6'' and provides simultaneous, AO-corrected imaging of the surroundings of the targets. Our preliminary results are :

\vspace*{2mm} \hspace*{.3cm} - \textbf{Multiplicity fraction :} 20-25\%\ of our targets has a very close companion detected by SAM (Fig.~1). Most of these detections are new.  This fraction increases to over 50\% if we include the wider pairs seen in the NACO field of view. Adding the results from other high-angular resolution imaging techniques (speckle, lucky imaging, AO) and from spectroscopy, only nine stars have no companion at all (among which two are known runaways).

\vspace*{2mm} \hspace*{.3cm} - \textbf{Separation distribution :} Fig.~2 (left panel) shows the cumulative number distribution of the measured separations in the range 30-6,000~mas and $\Delta mag<5$. The distribution is clearly double-peaked with an overabundance of pairs between 30 and 100~mas and between 1'' and 6''. 

\vspace*{2mm} \hspace*{.3cm} - \textbf{Brightness ratio distribution :} In these two separation ranges, the $\Delta mag$ distributions show different properties (Fig.~2, right panel), as confirmed by a Kolmogorov-Smirnov test at the 0.01 significance level. The wide pairs are dominated by fainter (most likely lower mass) companions while the distribution is almost uniform for the very close pairs. This suggests that the two populations have different natures.

\begin{figure}
  \includegraphics[width=11cm,height=7.5cm]{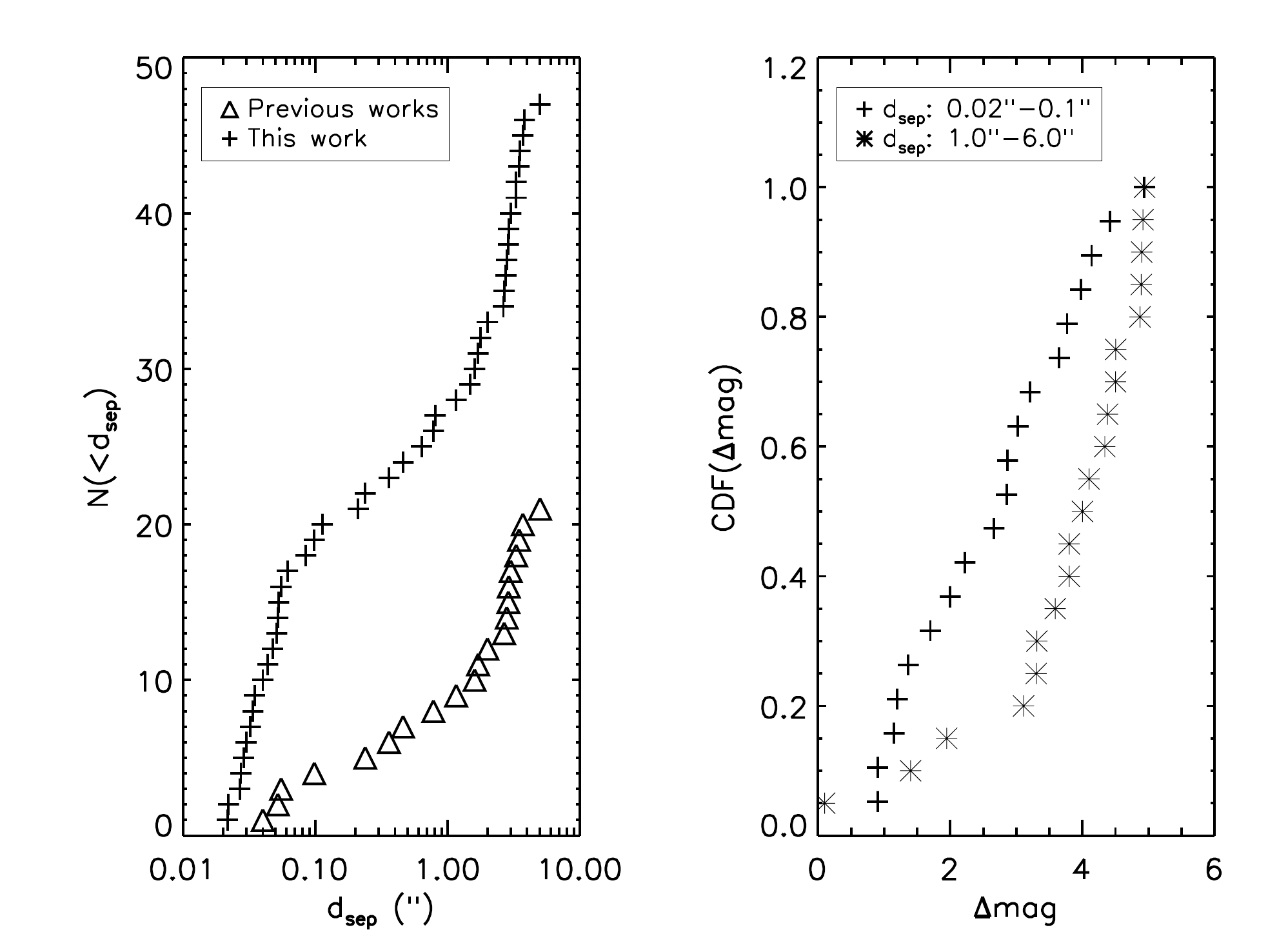}
  \caption{Left panel: Cumulative number distribution
of the separations up to 6\arcsec. The distribution shows two preferred separation ranges : from 30 to 100~mas and from 1'' to 6''. Right panel: Cumulative distribution function (CDF) of the magnitude differences in the two preferred separation. Only pairs with $\Delta\mathrm{mag}<5$ are considered to limit the
impact of the observational biases.} \label{fig2}
\end{figure}

\section{Perspectives and conclusions}
The SAM mode at NACO/VLT offers a new observing window  to study massive binaries allowing us to probe efficiently the short angular separation/high contrast regime. Future work involves (i) converting observational parameters to physical quantities, and (ii) investigating whether observational biases can explain the lack of companions in the 0.1-1.0” range. 

\acknowledgements 
Based on observations collected at the European Southern Observatory (Paranal, Chile) under program ID 086.D-0641.
\bibliography{hsana_sam}

\end{document}